\documentclass[12pt, prd, showpacs,superscriptaddress,notitlepage]{revtex4}
\usepackage[a4paper, margin=0.8in]{geometry}
\usepackage{setspace}

\usepackage{amssymb}
\usepackage{amsmath}
\usepackage[hidelinks=true]{hyperref}

\begin{document}

\title{Dirty rotating black holes:\\
regularity conditions on stationary horizons}
\author{I. V. Tanatarov}
\email{igor.tanatarov@gmail.com}
\affiliation{Kharkov Intitute of Physics and Technology,\\
1 Akademicheskaya, Kharkov 61108, Ukraine}
\affiliation{Department of Physics and Technology, Kharkov V.N. Karazin National
University, 4 Svoboda Square, Kharkov 61022, Ukraine}
\author{O. B. Zaslavskii}
\email{zaslav@ukr.net}
\affiliation{Department of Physics and Technology, Kharkov V.N. Karazin National
University, 4 Svoboda Square, Kharkov 61022, Ukraine}

\begin{abstract}
We consider generic, or ``dirty'' (surrounded by matter), stationary
rotating black holes with axial symmetry. The restrictions are found on the
asymptotic form of metric in the vicinity of non-extremal, extremal and
ultra-extremal horizons, imposed by the conditions of regularity of
increasing strength: boundedness on the horizon of the Ricci scalar, of
scalar quadratic curvature invariants, and of the components of the
curvature tensor in the tetrad attached to a falling observer.

We show, in particular, that boundedness of the Ricci scalar implies the
``rigidity'' of the horizon's rotation in all cases, while the finiteness of
quadratic invariants leads to the constancy of the surface gravity. We
discuss the role of quasiglobal coordinate $r$ that is emphasized by the
conditions of regularity. Further restrictions on the metric are formulated
in terms of subsequent coefficients of expansion of metric functions by $r$.
The boundedness of the tetrad components of curvature tensor for an observer
crossing the horizon is shown to lead in the horizon limit to
diagonalization of Einstein tensor in the frame of zero angular momentum
observer on a circular orbit (ZAMO frame) for horizons of all degrees of
extremality.
\end{abstract}

\maketitle

\newpage
\begin{spacing}{1.2}
\tableofcontents
\end{spacing}




\section{Introduction}

The properties of the near-horizon geometry play an essential role in many
important issues in gravitational physics. One of them is the constancy of
the surface gravity $\kappa $ that is crucial for the validity of the laws
of black hole mechanics \cite{bch} which were later recognized as laws of
thermodynamics with $\kappa $ related to the temperature of the system \cite%
{ev}. Another property is the constancy of the angular velocity of a black
hole $\omega _{H}$ that coincides with the horizon value of the metric
function $\omega $ responsible for rotation (see details below) and allows
one to consider a horizon as a solid rigidly rotating object. The
fundamental properties of the near-horizon geometry established in \cite{bch}
claim the constancy of $\kappa $ and $\omega _{H}$ but tell nothing of
the asymptotic form of the metric and the rate with which $\omega -\omega
_{H}$ approaches zero. Meanwhile, these details become very important, for
example, in the relation between the symmetries of the near-horizon geometry
and the universality of the Bekenstein-Hawking entropy. This served as a
motivation for closer examination of the properties of the metric near the
horizon for the \textquotedblleft dirty\textquotedblright\ (surrounded by
matter) black holes, undertaken in \cite{v1} for static and in \cite{v2} for
stationary geometries. The aforementioned papers were mainly devoted to the
nonextremal case. The ultraextremal one was only touched upon partially,
while the extremal case escaped consideration at all. Roughly speaking, the
extremal case corresponds to the degenerate horizons of the second order
while the ultraextremal case implies the multiple horizon of the order three
or higher. More explicit definition of such horizons in terms of the metric
functions will be given in the text below.

In the present paper, we extend the results of \cite{v2} to the extremal and
ultraextremal cases for which $\kappa =0$. We find the asymptotic expansion
of the metric near the horizon compatible with the regularity of the
geometry. It turns out that one cannot simply put $\kappa =0$ in the
formulas derived in \cite{v2} since the asymptotic expansion for $\omega $
changes qualitatively. It is worth stressing that knowledge of such
properties is necessary in a number of different physical problems. First of
all, it concerns the issue of the black hole entropy \cite{v1}, \cite{v2}
which in the (ultra)extremal case becomes much more subtle and even
contradictory \cite{eent}. The attempts to give a self-consistent
description of the entropy in the extremal case were made for the
spherically symmetric case \cite{isr}, \cite{plb} but to generalize them to
rotating geometries, detailed information on the asymptotic behavior of the
metric near the horizon is needed.

Apart from this, it is required by a number of concrete physical and
astrophysical applications. Let us give a couple of examples. Quite
recently, the effect of acceleration of particles by rotating black holes
(the BSW effect) was discovered for the Kerr extremal metric \cite{ban}.
Later on, it was generalized to generic rotating dirty black holes, both for
the extremal and nonextremal cases \cite{prd}. One of the key features of
this work consisted in a proper account for the difference in the asymptotic
behavior of the metric function $\omega $ responsible for rotation for extremal and nonextremal black holes (see eqs. (12) and (13) of \cite{prd}).
The validity of such an expansion in the extremal case can be checked for
the Kerr metric directly but it is important to have it in the general case
as well. Another example is studying the limiting transition from the rotating so-called quasiblack hole -- an object on the threshold of forming a horizon in a horizonless configuration -- to the black hole limit \cite{mein}.
Such a limit turns out to be well-defined only due to the validity of the same asymptotic expansion of $\omega $ (see eq.(5) in \cite{jqbh}). Apart from this, the properties of different kinds of observers in the vicinity of a black hole (orbiting around it or falling through the horizon) are important in the membrane formalism \cite{m}.

We find the restrictions which come from the two types of requirements. The
first one is the finiteness of some curvature invariants imposed on the
properties of the near-horizon geometry. This is done by examining the same
invariants that were considered in \cite{v1}, \cite{v2} for the nonextremal
case. Thus our approach is purely geometric and, similarly to \cite{v1}, 
\cite{v2} and contrary to \cite{bch}, does not use any energy conditions
imposed on matter near the horizon.

Meanwhile, there is also another regularity requirement not considered in 
\cite{v1}, \cite{v2}. It states that in the frame attached to an observer
crossing the horizon all curvature tensor components should be finite. If we
want the vicinity of the horizon to be regular, the requirement under
discussion imposes strong restrictions on the metric functions. If this
requirement is violated, while the first one from the previous paragraph is
satisfied, the corresponding space-time represents a so-called truly naked black hole (TNBH). Such objects, introduced and discussed in \cite{vo}, \cite{tr} and \cite{prd08}, generalized some previous observations made in \cite{nk1}, \cite{nk2}. From the mathematical viewpoint, their horizons are examples of the so-called nonscalar curvature singularities \cite{hawe} (Chap. 8), \cite{el}.

The paper is organized as follows. In Sec. II, we give basic definitions of
different types of horizons (nonextremal, extremal, ultraextremal) and
describe briefly the metrics we consider. In Sec. III, we examine the Ricci
scalar and the invariants quadratic with respect to the Ricci tensor and
derive the restrictions on the metric functions that are necessary and
sufficient for their finiteness. Also, we discuss the role of the
requirement of the metric's analyticity near the horizon and that of the quasiglobal coordinate. In Sec. IV, the
properties of the components of the Riemann tensor in the vicinity of the horizon are discussed for observers with zero angular momentum on circular orbits (OZAMOs) and for observers crossing the horizon. In sec. V we illustrate the general results on the example of Kerr-Newman vacuum solution. The on-horizon structure of the Einstein tensor in the OZAMO frame is considered in Sec. VI. Summary and Conclusion are given in Sec. VII. In Appendix (Sec. VIII) we give some useful formulas for the Einstein tensor in the OZAMO frame.

\section{Basic formulas}

\subsection{Types of horizons}

In the simplest, spherically symmetric case of a black hole metric in terms
of the so-called quasiglobal coordinate $r$ 
\begin{equation}
ds^{2}=-f(r)dt^{2}+\frac{dr^{2}}{f(r)}+R^{2}(r)d\Omega ^{2},  \label{f}
\end{equation}%
the event horizon is the surface $r=r_{h}=const$, on which $f(r)\rightarrow
0 $. We assume that near the horizon $f\sim (r-r_{h})^{p}$. For the horizon
to be non-singular and allow analytic continuation into the inner region, $p$
must be integer (for more on this see \cite{prd08}). The horizon is referred
to as non-extremal if $p=1$ (so $r=r_{h}$ is a simple zero of $f(r)$),
extremal if $p=2$, and ultraextremal if $p\geq 3$.

Making a change of variables from $r$ to the proper distance to the horizon $%
n$ and using the Gaussian normal coordinates we can write the metric in the
neighborhood of the horizon as%
\begin{equation}
ds^{2}=-N^{2}(n)dt^{2}+dn^{2}+r^{2}(n)d\Omega ^{2},  \label{sN}
\end{equation}%
where $N(n)$ is the lapse function. In the non-extremal case 
\begin{equation}
N(n)\sim \kappa n  \label{non}
\end{equation}%
at the horizon, where $\kappa $ is the surface gravity. In the extremal case 
$n\sim \ln (r-r_{h})$ and 
\begin{equation}
N\sim e^{-an}  \label{ext}
\end{equation}%
($a>0$ is a constant), and in the ultraextremal one 
\begin{equation}
N\sim n^{-k}\text{,}  \label{u}
\end{equation}%
with $k=\frac{p}{p-2}>0$. The condition $N=0$ on the horizon means that in
the extremal and ultraextremal cases it is situated at infinite distance $%
n\rightarrow +\infty $ .

In the absence of specific symmetries, the form of metric (\ref{f}) is no longer valid. Then, the asymptotic dependence of the lapse function on
the proper distance to the horizon can be taken as definition of the horizon
type: it is non-extremal if $N(n)\sim n$, extremal if (\ref{ext}) holds, and
ultraextremal in case relation (\ref{u}) holds with $k>0$. 

\subsection{Axially symmetric space-times}

Consider a generic axially symmetric rotating black hole space-time. Its metric in the vicinity of the horizon can be written in terms of Gaussian normal coordinates as 
\begin{equation}
ds^{2}=-N^{2}dt^{2}+g_{\phi \phi }(d\phi -\omega dt)^{2}+dn^{2}+g_{zz}dz^{2},
\label{z}
\end{equation}
where $n$ is the proper distance to the horizon, on which $N=0$. Due to the
symmetries, the metric functions here do not depend on $t$ and $\phi $.
Instead of $n$ and $z$ one can use coordinates $r$ and $\theta $, similar to
the Boyer-Lindquist ones for the Kerr metric. Throughout the paper we assume
that the fundamental constants $G=c=\hbar =1$.

In what follows, we use notations for the indices 
\begin{align}
\mu ,\nu ,\lambda ,\ldots =& 0,1,2,3=t,\phi ,n,z;  \label{indexes} \\
i,j,k,l,\ldots =& \quad 1,2,3=\quad \phi ,n,z; \\
A,B,C,\ldots =& 0,1\quad \quad =t,\phi ; \\
a,b,c,\ldots =& \quad \quad 2,3=\quad \quad n,z.
\end{align}%
Under the listed assumption of symmetry, the nonvanishing Christoffel
symbols for the metric (\ref{z}) read 
\begin{align}
& \Gamma _{A,aB}=\tfrac{1}{2}\partial _{a}g_{AB},\phantom{g^{AD}}\quad
\Gamma _{a,AB}\!=-\tfrac{1}{2}\partial _{a}g_{AB},\phantom{g^{ab}}\!\quad
\Gamma _{a,bc}={}^{(2)}\Gamma _{a,bc}, \\
& {\Gamma ^{A}}_{aB}=\tfrac{1}{2}g^{AD}\partial _{a}g_{BD},\quad {\Gamma ^{a}%
}_{AB}=-\tfrac{1}{2}g^{ab}\partial _{b}g_{AB},\quad {\Gamma ^{a}}%
_{bc}={}^{(2)}{\Gamma ^{a}}_{bc},
\end{align}%
where ${}^{(2)}\Gamma _{a,bc}$ and ${}^{(2)}{\Gamma ^{a}}_{bc}$ are the
Christoffel symbols for the two-dimensional metric $g_{ab}$.

\section{Curvature invariants}

\subsection{Ricci scalar}

In what follows, we need the expression for the Ricci scalar $R$. Let us
consider the foliation of the space-time by hypersurfaces $t=const$ with the
unit normal vectors $u_{\mu }=N\delta _{\mu }^{0}$. Then one can use the
general formula (see, e.g., eq. (3.43) in the textbook \cite{erik}) 
\begin{equation}
R={}^{(3)}\!R+(K^{ij}K_{ij}-K^{2})-2\chi _{;\mu }^{\mu }.
\end{equation}
Here semicolon denotes covariant derivative with respect to the metric $%
g_{\mu\nu}$; $^{(3)}\!R$ is the Ricci scalar of the slice $t=const$, $K_{ij}$
is its extrinsic curvature tensor 
\begin{equation}
K_{ij}=u_{\mu ;\nu }e_{(i)}^{\mu }e_{(j)}^{\nu },
\end{equation}%
where $e_{(i)}^{\mu }$ are the orthonormal basis vectors within the
hypersurface, and 
\begin{equation}
\chi ^{\nu }\equiv u^{\mu }u_{;\mu }^{\nu }-u^{\nu }u_{;\mu }^{\mu }.
\end{equation}

It is convenient to choose $e_{(i)}^{\mu }$ along the coordinate axes $(\phi
,n,z)$. Then, after straightforward calculations, we obtain that 
\begin{align}  \label{ExtrCurv}
\chi ^{0}=\chi ^{\phi }=0,& \quad \chi ^{a}=g^{ab}\partial _{b}\ln N, \\
K_{\phi \phi }=K_{ab}=0,& \quad K_{a\phi }=-\frac{g_{\phi \phi }}{2N}%
\,\partial _{a}\omega .
\end{align}%
Rewriting $\chi _{;\mu }^{\mu }$ in terms of the three-metric $g_{ik}$, we can recast the scalar curvature in the form 
\begin{equation}  \label{R}
R={}^{(3)}R +\frac{g_{\phi \phi }}{2N^{2}}(\nabla \omega )^{2} -2\frac{%
{}^{(3)}\!\Delta N}{N},
\end{equation}%
where $^{(3)}\!\Delta N$ is the Laplacian calculated with respect to the
metric $g_{ik}$ of the slice $t=const$; $(\nabla \omega
)^{2}=g^{ab}(\partial _{a}\omega )(\partial _{b}\omega )$ coincides for the
metrics $g_{\mu \nu }$, $g_{ij}$ and $g_{ab}$, since $\omega $ does not
depend on $t$ and $\phi $.

The third term comes from $\chi^{\mu}_{;\mu}$ and is the same as in the
static case \cite{v1}, while the second one appears due to non-zero
extrinsic curvature of hypersurfaces $dt=0$ in axially symmetric space-times.

The right hand part of (\ref{R}) in an explicit form is 
\begin{align}
R=& {}^{(3)}R+\frac{g_{\phi \phi }}{2N^{2}}\Big[(\partial _{n}\omega
)^{2}+g_{zz}^{-1}(\partial _{z}\omega )^{2}\Big]-  \notag \\
& -\Big\{2\Big(\frac{\partial _{n}^{2}N}{N}+g_{zz}^{-1}\frac{\partial
_{z}^{2}N}{N}\Big) +\partial _{n}(\gamma _{\phi }+\gamma _{z}) \cdot \frac{%
\partial _{n}N}{N} +g_{zz}^{-1}\partial _{z}(\gamma _{\phi } -\gamma
_{z})\cdot \frac{\partial_{z}N}{N}\Big\},  \label{Rcomponents}
\end{align}
where $\gamma _{\phi }=\ln g_{\phi \phi }$ and $\gamma _{z}=\ln g_{zz}$; the
terms in brackets come from extrinsic curvature, the ones in braces from $\chi_{\mu }^{;\mu }$.

\subsection{Non-extremal horizons}

In this section we repeat the results obtained in \cite{v2} in order to lay
down the scheme to be used later for (ultra)extremal horizons, write out the
results explicitly for meaningful comparison and interpretation, and correct
some minor errors made in the cited paper.

\subsubsection{Ricci scalar}

Let us consider $R$ in the vicinity of the horizon where, by definition, the
lapse function has the asymptotic form $N\sim \kappa n$ (\ref{non}). As all
derivatives of $N$ are bounded, the term $\sim {}^{(3)}\Delta N/N$ (the
terms in braces in Eq. (\ref{Rcomponents})) contains only divergences of the
order $\sim 1/n$. Meanwhile, assuming expansion of $\omega $ of the same type (we reserve notation $\omega_{i}$ to use it below)
\begin{equation}
\omega (n,z)=\omega _{H}(z)
	+\hat{\omega}_{1}(z)n
	+\hat{\omega}_{2}(z)n^{2}
	+\ldots ,  \label{omega-non0}
\end{equation}
the terms originating from the contribution of external curvature contain divergences $\sim ~\!1/n^{2}$: 
\begin{equation}
R=\frac{g_{\phi \phi }}{2}\Big[\Big(\frac{\partial _{n}\omega }{N}\Big)%
^{2}+g_{zz}^{-1}\Big(\frac{\partial _{z}\omega }{N}\Big )^{2}\Big]+O\Big(%
\frac{1}{n}\Big).
\end{equation}%
As it is a sum of squares, in order for them not to diverge stronger than
the remaining term $\sim 1/n$, we need both $\partial _{n}\omega =O(\sqrt{n})
$ and $\partial _{z}\omega =O(\sqrt{n})$, but taking into account that
expansion (\ref{omega-non0}) contains only integer powers of $n$, this is
reduced to 
\begin{equation}
\partial _{n}\omega =O(N),\qquad \partial _{z}\omega =O(N)  \label{OmegaNon}
\end{equation}%
when $N\rightarrow 0$. In terms of expansion coefficients, this means the
rigidity of the horizon's rotation $\omega _{H}=const$ and also that $\omega
_{1}=0$, so that expansion (\ref{omega-non0}) reads 
\begin{equation}
\omega (n,z)=\omega _{H}+\hat{\omega}_{2}(z)n^{2}
	+\hat{\omega}_{3}(z)n^{3}
	+\ldots .  \label{Omega-NonExp}
\end{equation}
It can also be rewritten as a series by $N$ 
\begin{equation}
\omega (n,z)=\omega _{H}+
	\tilde{\omega}_{2}(z)N^{2}
	+\tilde{\omega}_{3}(z)N^{3}
	+\ldots ,  \label{Omega-NonN}
\end{equation}
so $\partial_{z}\omega =O(N^2)$ also holds. Note that those are not
sufficient conditions for $R$ to be finite, but further constraints are more
conveniently obtained from the quadratic invariants.

\subsubsection{Quadratic invariants}

Let us now consider the traceless part $Q_{\mu\nu}=R_{\mu \nu }-\frac{1}{4}%
Rg_{\mu \nu }$ of the Ricci tensor squared 
\begin{equation}  \label{r2}
R_{2}\equiv 4Q_{\mu \nu }Q^{\mu \nu } \equiv 4R_{\mu \nu }R^{\mu \nu }-R^{2}.
\end{equation}

The expressions for the quadratic invariants do not seem to have an elegant
form in terms of the three- and two-dimensional geometries, and in their
explicit forms are rather lengthy, so we will not provide them here. In the
limit $n\rightarrow 0$ for the non-extremal horizon, however, one can obtain 
\begin{equation}
R_{2}=\Big[16\Big(\frac{\partial _{n}^{2}N}{\kappa }\Big)^{2}+\frac{8}{g_{zz}%
}\Big(\frac{\partial _{z}N}{N}\Big)^{2}+\big(\partial _{n}\gamma _{\phi }%
\big)^{2}+\big(\partial _{n}\gamma _{z}\big)^{2}\Big]_{H}\cdot \frac{1}{n^{2}%
}+O\Big(\frac{1}{n}\Big),  \label{R2-non}
\end{equation}%
where subscript $H$ denotes that all the quantities are taken at the
horizon. Note that the derivatives of $\gamma _{n,\phi }$ are by $n$, not by 
$z$ as given in \cite{v2}. This is again the sum of squares, and we obtain
four conditions on the metric functions, which lead to their expansions of
the form 
\begin{align}
N& =\kappa n+\hat{\kappa}_{3}(z)n^{3}+\ldots ;  \label{Non-N} \\
g_{\phi \phi }& =g_{\phi H}(z)+\hat{g}_{\phi 2}(z)n^{2}
	+\ldots ;
\label{Non-gphi} \\
g_{zz}& =g_{zH}(z)+\hat{g}_{z2}(z)n^{2}+\ldots,  \label{Non-gz}
\end{align}
where the surface gravity $\kappa $ is constant. Expansions (\ref{Omega-NonExp}) and (\ref{Non-N})--(\ref{Non-gz}) coincide with the
corresponding results of \cite{v2} (eqs. (15)--(17) and (19)). It is
straightforward to check that under those conditions the Ricci scalar (\ref{Rcomponents}) is regular. It turns out that the terms $\sim 1/n$ in $R_{2}$ also vanish, so both $R_{2}$ and the Kretchman scalar $Kr$ can be shown to
be bounded.

\subsection{(Ultra)extremal horizons}

\subsubsection{Extremal case}

In the neighborhood of an extremal horizon the lapse function has the
asymptotic form $N\sim e^{-an}$ (\ref{ext}) and we assume expansion of all
other metric functions in terms of $e^{-an}\to 0$ (the horizon is at $n\to
+\infty$). Then the term $\sim {}^{(3)}\Delta N /N$ in $R$ is regular, and
the only potentially dangerous terms that can diverge at the horizon in this
case are the ones in Eqs. (\ref{R}) and (\ref{Rcomponents}) with derivatives
of $\omega$, originating from the contribution of extrinsic curvature.
Therefore we are again led to 
\begin{equation}  \label{OmegaExtr}
\partial _{n}\omega =O(N), \qquad \partial _{z}\omega =O(N)
\end{equation}
for $N\rightarrow 0$, and the asymptotic expansion for $\omega $ reads 
\begin{equation}  \label{OmegaExtrExp}
\omega =\omega _{H}+\tilde{\omega}_{1}(z)N+O(N^{2}),
\end{equation}
where $\omega _{H}$ is constant. This is the same conclusion as for the
nonextremal case (\ref{Omega-NonN}), however, in contrast to nonextremal
horizons, now there is no restriction on $\tilde{\omega}_{1}$.

Looking into the quadratic invariants, on substitution of (\ref{OmegaExtrExp}%
) into (\ref{r2}) one can see that the invariant $R_{2}$ is finite on the
horizon. We have checked that under the same assumptions the Kretschmann
scalar is bounded as well, so the regularity of quadratic algebraic
invariants of the curvature tensor provides no additional constraints.

\subsubsection{An attempt to depart from analyticity}

For the metric (\ref{z}), $\kappa =\left( \frac{\partial N}{\partial n}%
\right) _{H}$. In what follows, we deal with the horizons for which $%
\kappa=0 $. One can try to extend the definition of the
extremal horizon and take, say, $a=a(z)$ in (\ref{ext}) instead of $a=const$. This choice, however, is incompatible with the requirement of analyticity
of the metric (\ref{f}).

Analyticity can be defined in terms of the quasiglobal coordinate $r$ which
behaves near the horizon as the Kruskal one \cite{k1}, \cite{k2}. This is
impossible for metric (\ref{z}) in the whole space-time, but nonetheless
such a coordinate can be introduced approximately in the vicinity of the
horizon: in the region where the Gaussian normal coordinates work, we can
always pass from variable $n$ to $r=r(n)$, such that metric has the form 
\begin{equation}  \label{axial-quasiglobal}
ds^{2}=-N^{2}(r,z)dt^{2} +g_{\phi \phi }(d\phi -\omega dt)^{2} +\frac{dr^{2}%
}{A(r)}+g_{zz}dz^{2},
\end{equation}
with $A(r)\sim N^{2}(r,z)\sim r^{p}$ for $N\to 0$. 

For nonextremal horizons ($p=1$) this was done in Sec. VII\ of Ref. \cite{inner} but generalization to multiple horizons ($p\geq 2$) is straightforward. The index $p$ should be integer if we want to have the
metric analytical and extendable across the horizon. This rules out the
dependence $a(z)$ in (\ref{ext}). Nonetheless, for completeness, we
investigate below the case of $a(z)$ as well, relaxing the requirement of
analyticity and relying on a weaker condition of the finiteness of curvature
invariants only.

We assume the asymptotic form for the lapse function in the vicinity of the
horizon $n\to+\infty$ is 
\begin{equation}
N=A(z)e^{-a(z)n}+B(z)e^{-2a(z)n}+\ldots \text{,}  \label{Nextr}
\end{equation}%
and examine whether such an asymptotic expression is compatible with the
finiteness of $R$ and $R_{2}$.


When $a=a(z)$, the term ${}^{(3)}\Delta N/N$ in $R$ (\ref{R}) is not
regular, but contains only polynomial divergences, and the exponential
divergences can only be in the term with the derivatives of $\omega $. The
conditions on the expansion of $\omega $, imposed in order to eliminate
them, are reduced then to the ones obtained in the simpler case (\ref%
{OmegaExtrExp}).

It turns out, that after substitution of (\ref{OmegaExtrExp}) and (\ref%
{Nextr}) into $R_{2}$, all the exponential divergences in $R_{2}$ vanish,
and only the polynomial ones remain. The worst possible remaining terms are
proportional to $n^{4}$, as each power of $n$ arises from differentiation by 
$z$, and it can be shown that 
\begin{equation}
R_{2}=\frac{n^{4}}{4}\Big[\frac{a^{\prime }(z)}{g_{\phi \phi }g_{zz}}\Big]%
_{H}^{4}\cdot \Big[\big(4-3g_{\phi \phi }\omega _{1}^{2}\big)^{2}+2g_{\phi
\phi }^{2}\omega _{1}^{4}\Big]_{H}+O(n^{3}).  \label{R2-n4}
\end{equation}
The factor by $n^{4}$ is an explicitly positive quantity, so the divergence
vanishes if and only if $a^{\prime }(z)=0$. Then all the lower degree
divergences also vanish and $R_{2}$ is regular. The Kretschmann scalar in
this case can be shown to be regular as well. Thus even the rather weak
requirement of finiteness of $R_{2}$ forces us to reject the dependence of $%
a $ on $z$.

\subsubsection{Quasiglobal coordinate}

Given two radial-type coordinates, $n$ and $r$, it is natural to ask the
question: in terms of which coordinate, $n$ or $r$ or some other, should we
write expansions of the metric functions?

Note now that restrictions on the metric we have obtained till now, by
ruling out scalar curvature singularities, are of two kinds: one is that
some expansion coefficients are zero and the other is that some other
expansion coefficients are constant on the horizon. The conditions of the
first type appear in the non-extremal case only.

For the sake of simplicity, from now on we redefine the coordinate $r$ in
such a way that $r_{h}=0$. For a non-extremal horizon $r\sim n^{2}$, so
expansion by $n$, from which \cite{v2} and we started, is a generalization
of expansion by $r$: if $g_{\mu \nu }$ is expanded into a series by $r$,
then $N(n)$ is a series with odd powers of $n$, while $\omega $, $%
g_{\phi\phi }$ and $g_{zz}$ are series with even powers of $n$. Thus the
obtained regularity conditions of the first type actually tell us that the first terms of the assumed expansions by $n$ are actually only integer powers of $r$: 
\begin{align}
g_{\mu \nu }(r,z)& =g_{\mu \nu }^{(h)}(z)+g_{\mu \nu }^{(1)}r+o(r);
\label{gmunuexp1} \\
N^{2}(r,z)& =\kappa ^{2}r+\tilde{\kappa}_{2}r^{2}+o(r^{2}).
\label{gmunuexp2}
\end{align}
It is likely that if we considered scalar curvature invariants involving
derivatives of higher order (e.g. $R_{\mu \nu \lambda \rho ;\sigma }R^{\mu
\nu \lambda \rho ;\sigma }$), we would obtain further conditions demanding
that terms with even powers of $n$ for $N$ and odd powers of $n$ for the
other metric functions should be zero.

The idea that metric functions should be expanded in terms of the quasiglobal coordinate is also strengthened by the analysis of extremal horizon made in the previous paragraph. In this case expansions of $\omega $ or $g_{ii}$ in terms of powers of $n$ are just not viable, leading inevitably to scalar curvature singularity, while expansions in terms of $r$ are perfectly admissible.

In section IV we will obtain further indications that this reasoning is correct from consideration of boundedness of tetrad components of the
curvature tensor in a tetrad attached to an observer crossing a non-extremal
horizon.


\subsubsection{Ultraextremal case}

For the ultra-extremal case expansions in terms of $n$ and in terms of $r$
are in general incompatible, as opposed to non-extremal case, because $r\sim
n^{m} $ with non-integer $m=-\frac{2}{p-2}$. In contrast to \cite{v2}, we
assume expansions of metric functions in terms of $r$: 
\begin{align}
N^{2}& =\kappa _{p}(z)r^{p}
+\kappa _{p+1}(z)r^{p+1}
+\kappa_{p+2}(z)r^{p+2}+O\left( r^{p+2}\right) ;  \label{Ultra-metric} \\
\omega & =\omega _{H}(z)+\omega _{1}(z)r
	+\omega_{2}(z)r^{2}+O(r^{3}); \\
g_{\phi \phi }& =g_{\phi H}(z)+g_{\phi 1}(z)r
	+g_{\phi 2}(z)r^{2}+O\left( r^{3}\right) ; 
\label{Ultra-metric1-5}\\
g_{zz}& =g_{zH}(z)+g_{z1}(z)r
	+g_{z2}(z)r^{2}+O\left( r^{3}\right) .
\label{Ultra-metric2}
\end{align}

Then the term $\sim {}^{(3)}\Delta N/N$ in $R$ (\ref{R}) is bounded, and in order to
have the Ricci curvature bounded, the two conditions 
\begin{equation}
\partial _{n}\omega =O(N),\qquad \partial _{z}\omega =O(N)
\label{Ultra-omega-gen}
\end{equation}%
must hold. The first one is satisfied automatically as soon as we assume expansions in terms of $r$: 
\begin{equation}
\frac{\partial _{n}\omega }{N}=\frac{dr}{dn}\frac{\partial _{r}\omega }{N}=%
\frac{\sqrt{A}}{N}\partial _{r}\omega \sim \partial _{r}\omega =O(1).
\end{equation}%
The second one implies that the first $(q+1)$ expansion coefficients of $%
\omega$, starting from $\omega_{H}$ and ending with $\omega_{q}$ with $%
q=[(p+1)/2]$ (brackets here denote integer part), do not depend on $z$, and
depending on the parity of $p$ the corresponding term in $R$ is either $O(1)$
or $o(1)$: 
\begin{align}
& \quad \omega =\omega _{H}+\ldots 
	+\omega_{q-1}r^{q-1}
	+\omega _{q}(z)r^{q}+\ldots ;  \label{ultra-omega} \\
p=2q-1:& \quad N\sim r^{q-1/2}\quad \Rightarrow \quad \partial _{z}\omega
\sim r^{q}=o(N); \\
p=2q\phantom{+1}:& \quad N\sim r^{q}\quad \Rightarrow \quad \partial
_{z}\omega \sim r^{q}=O(N).
\end{align}

The general form of expansions for quadratic invariants is rather
complicated, because the series starting with $r^{0}$ and ones starting with 
$r^{p}$ get intermixed. However, one can check that for specific cases $%
p=3,\ldots,8$, assuming the restrictions on $\omega$ obtained above hold,
both $R_{2}$ and $K$ are regular. Thus, just as in the extremal case, no
additional constraints are obtained.


\subsection{Intermediate results: eliminating scalar curvature singularity}

As mentioned above, the restrictions on metric obtained from the demand of
boundedness of scalar curvature invariants for any type of horizon are of
two kinds. Those of the first kind, when some expansion coefficients are
zero, can be summarized in the following way: expansion of the metric has the form (\ref{gmunuexp1},\ref{gmunuexp2}), i.e. the first several terms are a part of a series by the quasiglobal coordinate $r\sim n^{2}$ rather than by $n$.
This is obtained for non-extremal horizons, while for (ultra-)extremal ones
we assume expansion by $r$ from the very beginning, with $\kappa=0$ and the
first term $\sim r^{p}$.

The restrictions of the second kind are reduced to 1) restrictions on $%
\omega $ (\ref{Ultra-omega-gen}) 
\begin{equation}
\partial _{z}\omega =O(N)  \label{partZomega}
\end{equation}%
which enforce the ``rigidity'' of the horizon's rotation (in non-extremal
case this also implies $\partial_{z}\omega=O(N^2)$), and 2) the constancy of
the surface gravity (which holds by definition for (ultra)extremal horizons
with $\kappa =0$).

\section{Tetrad components of curvature tensor}

In the preceding section, we considered the conditions imposed on the metric by the finiteness of two curvature invariants. Meanwhile, more detailed information about the properties of the metric near the horizon is required. One may ask, what is the behavior of the gravitational characteristics which can be directly measured by an observer? In other words, we are interested in the components of the curvature tensor in the tetrad frame attached to different observers.

In particular, we consider two classes of them. The observer of the first type is orbiting a black hole outside of the horizon on a circular orbit, with its angular momentum equal to zero. Such observers were introduced in \cite{72} for the case of the Kerr metric and are usually referred to as ZAMOs (zero angular momentum observers), but we prefer to be more specific and will call them ``orbital zero angular momentum observers'', or OZAMOs for brevity. The observer of the second type falls through the horizon inside the black hole, freely or with finite proper acceleration, with conserving energy and angular momentum also equal to zero. We will call them ``falling zero angular momentum observers'', or FZAMOs for brevity.

Physically, the conditions on the metric which can be derived from the finiteness of the tetrad components of the curvature tensors can be
different for OZAMOs and FZAMOs. OZAMO frame is the most natural generalization of
the static frame in a static space-time to stationary metric. In the static
case, the Kretschmann invariant $Kr$ can be written in terms of the separate
components of the curvature tensor as a sum of squares, so the finiteness of 
$Kr$ requires the finiteness of each of them. Correspondingly, any algebraic
invariant composed from the curvature tensor will be also finite. In the
stationary case, the expression for $Kr$ includes terms with different signs
because of mixed components (with indices $0$ and $\phi $), so the general
picture is much more complicated.

Even in the static case, the components of the curvature tensor in the FZAMO
frame responsible for tidal forces can be significantly enhanced near the
horizon \cite{nk1}, \cite{nk2}. This amounts to infinite values of some of
them \cite{vo}, \cite{tr} which remains compatible with the finiteness of curvature invariants. The corresponding objects represent the so-called truly naked black holes. If we want the metric to be completely regular, we should
exclude such space-times. However, for this purpose, it is insufficient to examine the curvature invariants and the information about properties of FZAMO is required.

Thus we need to clarify the connection between (i) the finiteness of
curvature invariants and the conditions obtained in the OZAMO frame, (ii)
derive the corresponding conditions for FZAMO which are expected to give in
general additional constraints on the properties of the metric.

\subsection{Orbital ZAMO frame}

First of all, let us consider the components of curvature tensor $%
R_{\mu\nu\rho\sigma}$ in the coordinate frame $(t\phi nz)$. Using the Gauss-Codazzi equations \cite{erik} for the $R_{iklm}$ components (without zero
indices), we get 
\begin{equation}
R_{iklm}={}^{(3)}R_{iklm}+(K_{il}K_{km}-K_{ik}K_{lm}),
\end{equation}%
thus using (\ref{ExtrCurv}) we have explicitly 
\begin{align}
R_{abcd}& ={}^{(3)}R_{abcd}, \\
R_{a\phi b\phi }& ={}^{(3)}R_{a\phi b\phi }-\frac{g_{\phi \phi }^{2}}{4}%
\frac{\partial _{a}\omega }{N}\frac{\partial _{b}\omega }{N}, \label{Raphibphi}
\end{align}
while $R_{\phi abc}=0$ due to symmetry. Thus the conditions (\ref{OmegaNon})
are necessary for $R_{\mu\nu\rho\sigma}$ to be bounded, and direct
calculation of the other components shows that they are also sufficient for
that: 
\begin{equation}
R_{\mu\nu\rho\sigma}=O(1) \quad\Leftrightarrow\quad R=O(1).
\end{equation}
This fact does not seem to have any deep physical meaning by itself, though,
as the finiteness of curvature tensor in the given (badly behaved)
coordinate frame is irrelevant to regularity of geometry or observables.


Now let us consider an OZAMO -- zero angular momentum observer on a circular orbit $n=const$, $z=const$. The corresponding frame is defined by
the tetrad 
\begin{equation}
h_{(0)}=-Ndt,\quad h_{(1)} =\sqrt{g_{\phi \phi }}(d\phi -\omega dt),\quad
h_{(2)}=dn,\quad h_{(3)}=\sqrt{g_{zz}}\;dz.  \label{OZAMOframe}
\end{equation}
The tetrad components of the curvature tensor in this frame will be denoted
by overall tildes, i.e. $\tilde{R}_{\mu \nu \rho \sigma }$. As the
components in the coordinate frame $R_{\mu\nu\rho\sigma}$ are bounded and do
not generally tend to zero, and there is a divergent factor in $%
h_{(0)}^{\mu}\sim 1/N$, naive expectations would be that $\tilde{R}%
_{0i0j}\sim h_{(0)}^{\mu}h_{(0)}^{\nu}R_{\mu i\nu j}$ diverge as $1/N^2$ and 
$\tilde{R}_{0ijk}$ as $1/N$. However, as we show below, this is not the case.

The components $\tilde{R}_{a\phi b\phi }$ in the OZAMO frame differ from $R_{a\phi b\phi }$ (\ref{Raphibphi}) only by a factor of the order of unity, thus the conditions (\ref{OmegaNon}) are at least necessary for the boundedness of $%
\tilde{R}_{\mu \nu \rho \sigma }$. Assuming they hold, direct calculation
shows that all the other components of the curvature tensor in the OZAMO
frame are finite with the possible exception of 
\begin{align}
\tilde{R}_{0123}& =-\frac{1}{2}\sqrt{\frac{g_{\phi \phi }}{g_{zz}}}\;\frac{%
\partial _{z}\omega }{N}\frac{\partial _{n}N }{N}+O(1), \\
\tilde{R}_{0312}& =+\frac{1}{2}\sqrt{\frac{g_{\phi \phi }}{g_{zz}}}\;\frac{%
\partial _{z}\omega }{N}\frac{\partial _{n}N }{N}+O(1).
\end{align}%
However, as $\partial _{n}N/N\sim \partial _{r}N\sim r^{p/2-1}$, in the
(ultra-)extremal case $p\geq 2$ those are also regular, so no additional
constraints appear; and in the non-extremal case the two conditions (\ref%
{OmegaNon}) actually imply that $\partial_{z}\omega =O(N^2)$ (see (\ref%
{Omega-NonExp})), so there are no additional constraints in this case
either. Thus for all types of horizons 
\begin{equation}
\tilde{R}_{\mu\nu\rho\sigma}=O(1) \quad\Leftrightarrow\quad R=O(1).
\end{equation}
We see therefore that in the vicinity of a regular horizon an OZAMO observer
always experiences finite tidal forces, even though his acceleration itself
may diverge (at a non-extremal horizon).

\subsection{Falling ZAMO frame}

\subsubsection{Choice of tetrad for a falling observer}

If there's a Killing vector $\xi^\mu$, then the quantity $\xi^{\mu}u_{\mu}$
is conserved for a free-falling particle with $4$-velocity $u^{\mu}$.
Moreover, conservation of $u^{\mu}\xi_{\mu}$ along the worldline of any
particle, not necessarily free falling, is equivalent to its acceleration
along the Killing vector field being zero: 
\begin{equation}  \label{KillingProjection}
a^{\nu}\xi_{\nu}=(u^{\mu}\nabla_{\mu}u^{\nu})\xi_{\nu}
=u^{\mu}\nabla_{\mu}(u^{\nu}\xi_{\nu})- u^{\mu}u^{\nu}\nabla_{\mu}\xi_{\nu}
=u^{\mu}\nabla_{\mu}(u^{\nu}\xi_{\nu}).
\end{equation}
As the two Killing vectors for axially symmetric metric are $\partial_{t}$
and $\partial_{\phi}$, the two corresponding quantities that conserve on
geodesics are energy $E=-u_{t}$ and angular momentum $L=u_{\phi}$ (for
particles of unit mass, which will be considered hereafter).

Let us consider FZAMO, i.e. the observer with zero angular momentum and
conserving energy that is falling towards the horizon relative to an OZAMO.
According to what is written above, the $t$ and $\phi $ components of his
acceleration are zero. So, this class of observers obviously includes some
of the free-falling ones, with corresponding values of $E$ and $L$.

We attach the tetrad to FZAMOs and investigate the restrictions on the metric that would ensure that the curvature tensor components in this tetrad are bounded. The tetrad can be built in three steps:

\begin{itemize}
\item First, we take the tetrad of OZAMO frame (\ref{OZAMOframe}).

\item Second, we rotate the frame in the $n-z$ plane by angle $\theta$ 
\begin{equation}
\tilde{e}_{(2)}=h_{(2)}\cos\theta +h_{(3)}\sin\theta,\qquad \tilde{e}%
_{(3)}=-h_{(2)}\sin\theta +h_{(3)}\cos\theta.
\label{tetrad-mid}
\end{equation}

\item Finally, make a boost in the direction of $\tilde{e}_{(2)}$: 
\begin{equation}  \label{tetrad-fin}
e_{(0)}=\gamma \big(h_{(0)}+v \tilde{e}_{(2)}\big), \qquad e_{(1)}=h_{(1)},
\qquad e_{(2)}=\gamma \big(\tilde{e}_{(2)}+v h_{(0)}\big), \qquad e_{(3)}=%
\tilde{e}_{(3)},
\end{equation}
where 
\begin{equation}
\gamma=\frac{E}{N}
\end{equation}
is the Lorentz factor of the observer with respect to OZAMO frame, $v=\sqrt{%
1-1/\gamma^{2}}$ is the physical velocity, $v_{n}=v\cos\theta$ and $%
v_{z}=v\sin\theta$ are its spatial components.
\end{itemize}

We do not consider below particles with $E=0$. Those are a specific case of
critical particles, defined in general by relation $E-\omega_{H}L=0$, which
are special and give rise to many interesting phenomena, such as the BSW
effect \cite{ban}.

We will also impose one additional restriction on the observers: 
\begin{equation}
\theta =O(N).  \label{theta-small}
\end{equation}%
It always holds in the spherically symmetric case due to additional constant
of motion. In the axially symmetric non-extremal case it can be shown to
follow explicitly from the boundedness of a particle's scalar acceleration.
Extremal, and all the more so ultra-extremal general axially symmetric
metrics in principle allow motion even with $\theta =O(1)$. However, such
particles and corresponding metrics are also in a sense quite special,
unique to extremal horizons, and should be investigated separately. We will
not go into details here (they will be reported elsewhere), but instead,
though it may seem rather arbitrary, just assume (\ref{theta-small}) in all
cases.

\subsubsection{Non-extremal horizons}

The components of curvature tensor in the tetrad frame $\{e_{(i)}\}$ (\ref%
{tetrad-fin}), attached to FZAMO, will be denoted by overall hats $\hat{R}%
_{\mu \nu \rho \sigma }$. Assuming expansions (\ref{Omega-NonExp},\ref{Non-N}%
,\ref{Non-gphi},\ref{Non-gz}), the components responsible for the tidal
forces experienced by the observer are as follows near the horizon: 
\begin{align}
\hat{R}_{0101}& =-\frac{3E^{2}}{2\kappa^2 g_{\phi H}}
	\;	\frac{\hat{g}_{\phi 3}}{n}+O(1); \\
\hat{R}_{0202}& =O(1); \\
\hat{R}_{0303}& =-\frac{3E^{2}}{2\kappa^2 g_{\phi H}}\;
	\frac{\hat{g}_{z3}}{n}+O(1);\\
\hat{R}_{0102}& =-\frac{3E}{2\kappa^2}
	\sqrt{g_{\phi H}}\;
	\;\frac{\hat{\omega}_{3}}{n}+O(1); \\
\hat{R}_{0103}& =\frac{E^{2}\sqrt{g_{\phi H}}}{2\kappa ^{3}\sqrt{g_{zH}}}
\Big(\frac{\theta }{n}\;3\sqrt{g_{zH}}\;
	\hat{\omega}_{3}-\hat{\omega}_{3}^{\prime}
	\Big)\frac{1}{n}+O(1); \\
\hat{R}_{0203}& =O(1).
\end{align}
The necessary and sufficient conditions for them to be bounded are 
\begin{equation*}
\hat{g}_{z3}=\hat{g}_{\phi 3}=\hat{\omega}_{3}=0.
\end{equation*}
It can be checked that all the other components of curvature tensor are also bounded under these assumptions.

Together with conditions (\ref{gmunuexp1},\ref{gmunuexp2}), already taken
into account, these restrictions can be reformulated especially simply in terms of quasiglobal coordinate: 
\begin{equation}
g_{\mu \nu }=g_{\mu \nu }^{(h)}+g_{\mu \nu }^{(1)}r+g_{\mu \nu
}^{(2)}r^{2}+o(r^{2}).
\end{equation}%
Thus when we excluded scalar curvature singularity, we had to assume that metric is expanded in terms of $r$ instead of $n$ up to at least the terms $%
\sim r\sim n^{2}$ (\ref{gmunuexp1}), with $N^{2}$ up to $\sim r^{2}\sim n^{4}$ (\ref{gmunuexp2}), and now, excluding also truly naked black holes, we have to demand that the same condition holds for all components of $g_{\mu \nu }$ at least up
to terms $\sim r^{2}\sim n^{4}$.

On the other hand, we see that there is no restriction on the first non-zero expansion coefficients of $g_{\mu\nu}$ after the leading terms -- they are allowed to be functions of z -- and in this sense the restrictions obtained from the OZAMO and FZAMO frames are the same. We will see below that in the (ultra)extremal case this is not so.

\subsubsection{Extremal horizons}

In order to extract the necessary and sufficient conditions of regularity of
the curvature tensor in the FZAMO tetrad in the extremal and ultra-extremal
cases it appears not to be enough to consider only the tidal forces $\hat{R}%
_{0i0j}$ ($i,j=1,2,3$). So we look at all the components in the order that
allows us to extract those conditions most efficiently.

Assuming the only restriction obtained for extremal horizons from the demand
of curvature invariants' regularity $\omega _{H}=const$ (\ref{OmegaExtrExp}), the smallness of $\theta $ (\ref{theta-small}) and expansion of metric in terms of $r$ of general form (\ref{Ultra-metric})--(\ref{Ultra-metric2}) for extremal horizons, with $p=2$, so that 
\begin{equation}
N^{2}=\kappa _{2}(z)r^{2}+\kappa _{3}(z)r^{3}+O(r^{4}),
\end{equation}
the asymptotics of some of the tetrad components are given by 
\begin{align}
\hat{R}_{0101}\sim \hat{R}_{1212}\sim \hat{R}_{0112}& \sim \frac{1}{r^{2}} %
\big[g_{\phi H}^{\prime }\kappa _{2}^{\prime }+g_{\phi H}^{2}(\omega
_{1}^{\prime })^{2}\big]; \\
\hat{R}_{0103}\sim \hat{R}_{0123}\sim r\hat{R}_{0312}\sim r\hat{R}_{1223}&
\sim \frac{1}{r^{2}}\big[2\kappa _{2}\omega _{1}^{\prime }-\omega _{1}\kappa
_{2}^{\prime }\big]; \\
\hat{R}_{0203}\sim \hat{R}_{0223}& \sim \frac{1}{r}\big[2\kappa _{2}^{\prime
}-3g_{\phi H}\omega _{1}\omega _{1}^{\prime }\big]; \\
\hat{R}_{0113}& \sim \frac{\omega _{1}\omega _{1}^{\prime }}{r}.
\end{align}%
For their boundedness it is necessary that $\omega _{1}^{\prime }=\kappa
_{2}^{\prime }=0$. Under those conditions the expressions for $\hat{R}_{\mu \nu \rho \sigma
}$ are further simplified and we see, in particular, that 
\begin{equation}
\hat{R}_{0101}\sim \hat{R}_{0103}\sim \hat{R}_{1212}\sim \hat{R}_{0112}\sim 
\frac{\kappa _{3}^{\prime }}{r}.
\end{equation}%
Thus the necessary conditions are 
\begin{equation}
\omega _{1}^{\prime }=\kappa _{2}^{\prime }=\kappa _{3}^{\prime }=0.
\label{Extr-TNBH}
\end{equation}%
They can be rewritten as 
\begin{equation}
\partial _{z}\omega =O(N^{2}),\qquad \frac{\partial _{z}N^{2}}{N^{2}}%
=O(N^{2}).  \label{Non-Extra-Gen}
\end{equation}%
This set now can be verified to be sufficient for all the tetrad components
of the curvature tensor to be bounded. They are clearly more strict
conditions on the metric than just those that are necessary for the
boundedness of scalar invariants (\ref{OmegaExtr}).

\subsubsection{Ultra-extremal horizons}

In this case, it is more convenient to work in terms of metric functions and their asymptotics, rather than in terms of their expansions, as the series
starting with $r^{0}$ and $r^{p}$ are multiplied and divided, so the
resulting series for arbitrary $p$ are hard to deal with. Thus we do the
following. First, we calculate the exact expressions for the components of
the curvature tensor in the FZAMO tetrad, and inspect them for divergences
in assumptions (\ref{theta-small}) and 
\begin{equation}
N^{2}\sim A(r)\sim r^{p},\qquad p>2,\qquad \partial _{z}\omega =O(N),
\end{equation}%
where the last condition comes from the finiteness of curvature scalars according to (\ref{partZomega}). Then starting from some component we write
out the explicit condition of its boundedness in the form $f(g_{\mu \nu
},\partial _{\lambda }g_{\mu \nu })=O(1)$, where $f$ includes all the
potentially divergent terms. This condition is then used to simplify the
expressions for the other components, and allows us to write out the next
condition that is needed for the finiteness of the next component. This
procedure is repeated until all the components are finite, and this gives
us, by construction, the set of \emph{necessary and sufficient} conditions
for $\hat{R}_{\mu \nu \rho \sigma }$ to be regular. The choice of specific
succession can significantly simplify the conditions and their subsequent
reduction. Our choice leads to the following sequence: 
\begin{align}
1.\quad \hat{R}_{0113}:& \quad \partial _{r}\omega \cdot \partial _{z}\omega
=O(N^{2});  \label{Ultra-1} \\
2.\quad \hat{R}_{0101}:& \quad \partial _{z}\gamma _{\phi}\frac{\partial
_{z}N^{2}}{N^{2}}+g_{\phi \phi }\Big(\frac{\partial _{z}\omega }{N}\Big)%
^{2}=O(N^{2});  \label{Ultra-2} \\
3.\quad \hat{R}_{0203}:& \quad \frac{\partial _{z}N^{2}}{N^{2}}=O(r)\quad
\Leftrightarrow \quad \kappa _{p}^{\prime }=0;  \label{Ultra-3} \\
4.\quad \hat{R}_{0303}:& \quad 
\partial _{z}(\gamma _{z}-3\gamma _{\phi })
	\frac{\partial _{z}N^2}{N^2}
	+\Big(\frac{\partial_{z}N^2}{N^2}\Big)^{2}
	-2\frac{\partial _{z}^{2}N^2}{N^2}=O(N^{2});
\label{Ultra-4} \\
5.\quad \hat{R}_{0313}:& \quad \partial _{z}(\gamma _{z}-3\gamma _{\phi
})\cdot \partial _{z}\omega -2\partial _{z}^{2}\omega +\frac{\partial
_{z}N^{2}}{N^{2}}\partial _{z}\omega =O(N^{2});  \label{Ultra-5} \\
6.\quad \hat{R}_{0103}:& \quad \partial _{r}(\gamma _{z}-3\gamma _{\phi
})\cdot \partial _{z}\omega -2\partial _{r}\partial _{z}\omega +2\frac{%
\partial _{r}N^{2}}{N^{2}}\partial _{z}\omega -\frac{\partial _{z}N^{2}}{%
N^{2}}\partial _{r}\omega =O(N^{2}).  \label{Ultra-6}
\end{align}

Note that if (\ref{Non-Extra-Gen}) hold, then the six conditions (\ref{Ultra-1}--\ref{Ultra-6})
are also satisfied. So (\ref{Non-Extra-Gen}) are sufficient but not in
general necessary conditions for the ultra-extremal case.

It should be noted that the order of the obtained restrictions on metric is
not what would be expected from naive considerations: while the components
of curvature tensor in OZAMO frame are bounded, which is ensured by
conditions $\partial_{z}\omega=O(N)$ etc. (and which in itself is not
expected from naive considerations), making the boost into the FZAMO frame
with Lorentz factor $\sim 1/N$, we introduce divergent factors, the worst of
which are $\sim 1/N^2$; so one would expect that regularity conditions
change to e.g. $\partial_{z}\omega =O(N^3)$. The correct conditions turn out to be much softer, and the restriction (\ref{theta-small}) is essential here.

\subsubsection{Simplest case: $p=3$}

Let us obtain the explicit restriction on the metric that the six conditions (\ref{Ultra-1}--\ref{Ultra-6}) imply in terms of the metric functions' expansion coefficients by $r$ (\ref{Ultra-metric}--\ref{Ultra-metric2}) in the simplest case of ultra-extremal horizon, for $p=3$. Condition 2 is equivalent to $\kappa_{p}'=0$; after substitution of (\ref{Ultra-metric}--\ref{Ultra-metric2}) into the left hand side of the remaining five, we demand that all terms up to $\sim r^{2}$ are equal to zero, as in this case $O(N^{2})$
is $O(r^{3})$. After some algebra we obtain that there are three possible
variants.

\begin{enumerate}
\item The first is the one to be expected, with $\partial_{z}N^{2}=O(N^4)$
and $\partial_{z}\omega =O(N^2)$, which in terms of expansion coefficients
means 
\begin{align}
&\kappa_{3}^{\prime}=\kappa_{4}^{\prime}=\kappa_{5}^{\prime}=0; \\
&\omega_{H}^{\prime}=\omega_{1}^{\prime}=\omega_{2}^{\prime}=0
\end{align}
with no other constraints. The expansions themselves are then
\begin{align}
N^{2}(r,z)&=\underline{\kappa_{3}}r^{3} +\underline{\kappa_{4}}r^{4} +%
\underline{\kappa_{5}}r^{5} +\kappa_{6}(z)r^{6}+\ldots; \\
\omega(r,z)&=\underline{\omega_{H}} +\underline{\omega_{1}}r +\underline{
\omega_{2}}r^{2} +\omega_{3}(z)r^{3}+\ldots,
\end{align}
with $g_{\phi\phi}$ and $g_{zz}$ given by the general formulas (\ref{Ultra-metric1-5}), (\ref{Ultra-metric2}). Here and below in this section we underline the coefficients in the expansions that do not depend on $z$.

\item The second possible variant is more exotic: 
\begin{align}
&\kappa_{3}^{\prime}=\kappa_{4}^{\prime}=0, \quad\text{but}%
\quad\kappa_{5}^{\prime}\neq 0; \\
&\omega_{H}^{\prime}=\omega_{2}^{\prime}=0,\quad \omega_{1}=0; \\
&g_{\phi H}^{\prime}=0; \\
& g_{z H}=const\cdot (\kappa_{5}^{\prime})^{2},
\end{align}
so that the expansions (\ref{Ultra-metric}--\ref{Ultra-metric2}) read 
\begin{align}
N^{2}(r,z)&=\underline{\kappa_{3}}r^{3} +\underline{\kappa_{4}}r^{4}
+\kappa_{5}(z)r^{5}+\ldots; \\
\omega(r,z)&=\underline{\omega_{H}} +\underline{\omega_{2}}r^{2}
+\omega_{3}(z)r^{3}+\ldots; \\
g_{\phi\phi}(r,z)&=\underline{g_{\phi H}} 
	+g_{\phi 1}(z)r+\ldots; \\
g_{zz}(r,z)&
	=const\cdot(\kappa_{5}^{\prime})^{2}
	+g_{z1}(z)r+\ldots.
\end{align}

\item Lastly, there is one more variant, most exotic, with expansion coefficients 
\begin{align}
&\kappa_{3}^{\prime}=0,\quad\text{but}\quad \kappa_{4}^{\prime}\neq 0; \\
&\omega_{H}^{\prime}=0,\quad \omega_{1}=\omega_{2}=0; \\
&g_{\phi H}^{\prime}=g_{\phi 1}^{\prime}=0, \\
&g_{z H}=C_{1}\cdot (\kappa_{4}^{\prime})^{2}, \quad
g_{z 1}=g_{z H}\Big(
 	2\frac{\kappa_{5}'}{\kappa_{4}'}
	 -\frac{\kappa_4}{\kappa_3}+C_{2}	\Big),
\end{align}
where $C_{1,2}$ are constants, such that the metric functions are 
\begin{align}
N^{2}(r,z)&=\underline{\kappa_{3}}r^{3} +\kappa_{4}(z)r^{4}
+\kappa_{5}(z)r^{5}+\ldots; \\
\omega(r,z)&=\underline{\omega_{H}} +\omega_{3}(z)r^{3}+\omega_{4}(z)r^{4}+%
\ldots; \\
g_{\phi\phi}(r,z)&=\underline{g_{\phi H}} +\underline{g_{\phi 1}}r+g_{\phi 2}(z)r^{2}+\ldots;
\\
g_{zz}(r,z)&=C_{1}\cdot(\kappa_{4}^{\prime})^{2} \Big[1+\Big(
 	2\frac{\kappa_{5}'}{\kappa_{4}'}
	 -\frac{\kappa_4}{\kappa_3}+C_{2}\Big)r\Big]
 +g_{z 2}(z)r^{2}+\ldots.
\end{align}
\end{enumerate}

For larger values of $p$ we would obtain more exotic variants of regular
horizons.

\section{Example: Kerr-Newman metric}

To illustrate the above properties, let us consider the simplest and, at the
same time, one of the most physically important cases -- the Kerr-Newman
metric. In the Boyer-Lindquist coordinates this metric  can be written as \cite{gh77} 
\begin{align}
ds^{2} &=-\Big(1-\frac{2Mr-Q^{2}}
	{\Sigma }\Big)dt^{2}-\frac{2a(2Mr-Q^{2})
	\sin ^{2}\Theta }{\Sigma }dt\,d\phi 
	+\frac{\Sigma }{\Delta }dr^{2}
	+\Sigma d\Theta ^{2}+  \nonumber \\
&+\Big(r^{2}+a^{2}+\frac{(2Mr-Q^{2})a^{2}\sin ^{2}\Theta }{\Sigma }\Big)\sin ^{2}\Theta d\phi ^{2},
\end{align}%
where $a$ is the angular momentum parameter, $Q$ is the electric charge, $\Sigma \equiv r^{2}+a^{2}\cos ^{2}\Theta $, 
\begin{equation}
\Delta =r^{2}-2Mr+a^{2}+Q^{2}=(r-r_{+})(r-r_{-}),
\end{equation}
where $r_{\pm }=M\pm \sqrt{M^{2}-Q^{2}}$ are the roots of equation $\Delta =0$. The larger root $r_{+}$ corresponds to the event horizon.

The expression for the metric can be rewritten in the form (\ref{z}) with 
\begin{align}
\omega&=\frac{a(2Mr-Q^{2})}{\Sigma (r^{2}+a^{2})+(2Mr-Q^{2})a^{2}\sin
^{2}\Theta }=\frac{a(2Mr-Q^{2})}{(r^{2}+a^{2})^{2}-a^{2}\Delta\sin ^{2} \Theta },  \label{wnk} \\
N^{2}&=\frac{\Delta \Sigma } {(r^{2}+a^{2})^{2}
	-a^{2}\Delta\sin ^{2}\Theta }.  \label{nk}
\end{align}

Two cases should be considered separately.

\subsection{Non-extremal horizon, $M^{2}>a^{2}+Q^{2}$}

Let us consider the near-horizon region, $r\rightarrow r_{+}$. Then, the proper distance between the points with coordinates $r$ and $r_{+}$ is
\begin{equation}
n\approx 2\sqrt{\Sigma_{+}}\;
	 \frac{\sqrt{r-r_{+}}}{\sqrt{r_{+}-r_{-}}},
\end{equation}
where $\Sigma _{+}\equiv r_{+}^{2}+a^{2}\cos ^{2}\Theta $. Correspondingly, eq. (\ref{nk}) gives us 
\begin{equation}
N\approx \kappa n
\end{equation}
where 
\begin{equation}
\kappa =\frac{r_{+}-r_{-}}{2(r_{+}^{2}+a^{2})}
\end{equation}%
is the surface gravity of the Kerr-Newman black hole.\ It follows from (\ref%
{wnk}) that%
\begin{equation}
\omega -\omega _{H}\approx a(r-r_{+})\Big[\frac{2(M-2r_{+})} {%
(r_{+}^{2}+a^{2})^{2}}+\frac{a^{2}\sin ^{2}\Theta\; (r_{+}-r_{-})}{%
(r_{+}^{2}+a^{2})^{3}}\Big],  \label{ww}
\end{equation}%
where $\omega _{H}=\omega (r=r_{+})=a(r_{+}^{2}+a^{2})^{-1}=const$. Thus the
first correction to $\omega _{H}$ has the order $r-r_{+}\sim N^{2}$, in agreement with (\ref{Omega-NonN}).

\subsection{Extremal horizon, $M^{2}=a^{2}+Q^{2}$}

Then, $r_{+}=r_{-}$, $\Delta =(r-r_{+})^{2}$. The corrections to $\omega _{H}
$ have the order $r-r_{+}$ again. However, now $r-r_{+}\sim N$. Moreover,
the term with angular dependence in (\ref{ww}) vanishes, so $\omega
-\omega_{H}\approx \tilde{\omega}_{1}N$ in agreement with (\ref{OmegaExtrExp}), where $\tilde{\omega}_{1},\omega_{1}=const$ according to (\ref{Extr-TNBH}).

To obtain the triple horizon, one may introduce into consideration the cosmological term. However, we will not list the corresponding rather cumbersome expressions since the the triple root does not correspond in this case to the black hole horizon, as the region with $r>r_{+}$ is a cosmological one, with the positive factor in $ds^{2}$ at $dt^{2}$.

\section{On-horizon structure of the Einstein tensor}

The important property of the horizon consists in that the Einstein tensor in the OZAMO frame becomes diagonal in the horizon limit. If the Einstein equations are satisfied, this leads to important constraints on the possible form of the equation of state near the horizon. Diagonality of the Einstein tensor was demonstrated in \cite{v2} for non-extremal horizons by direct calculations in the tetrad that does not coincide with that of OZAMO and is not orthogonal, although the discrepancy becomes negligible in the limit under consideration. Meanwhile, the lack of orthogonality generates its own corrections which are hard to control.

Here, we show that the Einstein tensor indeed becomes diagonal in the orthogonal frame attached to an OZAMO. To this end, we use two different approaches. First, we calculate the components of this tensor directly like it was done in \cite{v2}, but for a different, exactly orthonormal, tetrad. Second, we show that the structure of the Einstein tensor can be understood if one takes into account the relationship between OZAMO and FZAMO.

\subsection{Direct calculation}

We want to establish the asymptotic structure of the Einstein tensor in the
OZAMO frame, denoted by tildes. The general expressions, though lengthy, are
quite manageable and are given in appendix. We are interested in the
difference 
\begin{align}
\tilde{G}^{0}_{0}-\tilde{G}^{2}_{2}=& \frac{g_{\phi\phi}}{2 g_{zz}} \Big(%
\frac{\partial_{z}\omega}{N}\Big)^{2} +\frac{1}{2}\Big\{D^{2}_{n}g_{zz}
+D^{2}_{n}g_{\phi\phi}\Big\}+  \notag \\
&-\frac{1}{2}\frac{\partial_n N}{N} \partial_{n}(\gamma_\phi +\gamma_z ) -%
\frac{1}{2g_{zz}}\frac{\partial_z N}{N} \partial_{z}(\gamma_\phi -\gamma_z )
-\frac{1}{g_{zz}}\frac{\partial_{z}^{2}N}{N}
\end{align}
and the two non-zero off-diagonal components $\tilde{G}^{0}_{1}$ and $\tilde{G}^{2}_{3}$ (for explicit expressions and definition of operator $D^{2}_{a}$
(\ref{D2}) see appendix; the others are zero due to symmetry). It was shown in \cite{v2} that in the horizon limit for the non-extremal case all the three quantities tend to zero, so that the Einstein tensor diagonalizes. Let us show now that the regularity conditions derived above imply that it actually diagonalizes for arbitrary horizons, either extremal or not, in a quite general manner.

We start from the $\tilde{G}^0_0 -\tilde{G}^2_2$ term. Rewriting the parts
with derivatives by $n$ in terms of $\partial_{r}=\sqrt{A}\;\partial_{n}$,
we get 
\begin{equation}
\tilde{G}^0_0 -\tilde{G}^2_2 = \frac{1}{2g_{zz}}\Big\{ g_{\phi\phi} \Big(%
\frac{\partial_z \omega}{N}\Big)^{2} -\frac{\partial_z N}{N}
\partial_{z}(\gamma_\phi -\gamma_z) -2\frac{\partial^2_z N}{N}\Big\} +\frac{%
A(r)}{2}\Big[D^{2}_{r}g_{\phi\phi} +D^{2}_{r}g_{zz} +\partial_{r}\ln\frac{A}{%
N^2}\Big].
\end{equation}
The second term here is $O(N^2)$, as $A\sim N^{2}$ and therefore $%
\partial_{r}\ln (A/N^2)=O(1)$, for arbitrary horizons.

\begin{itemize}
\item In the non-extremal and extremal cases the expression in the braces is
also $O(N^2)$ due to the regularity conditions in the form (\ref%
{Non-Extra-Gen});

\item In order to see that the same holds in the ultra-extremal case, we rewrite it as 
\begin{align}
\tilde{G}^0_0 -\tilde{G}^2_2 &= \frac{1}{2g_{zz}}\Big\{ g_{\phi\phi}\Big(\frac{%
\partial_z \omega}{N}\Big)^{2} +\partial_{z}\gamma_{\phi}\; \frac{%
\partial_{z}N^2}{N^2}\Big\}+  \notag \\
&+\frac{1}{4g_{zz}}\Big\{ 
	\partial _{z}(\gamma _{z}-3\gamma _{\phi })
	\frac{\partial _{z}N^2}{N^2}
	+\Big(\frac{\partial_{z}N^2}{N^2}\Big)^{2}
	-2\frac{\partial _{z}^{2}N^2}{N^2}\Big\}+ O(N^2).
\end{align}
The expression in the first braces is $O(N^2)$ due to condition 2 (\ref{Ultra-2}), and the one in the second braces is also $O(N^2)$ due to
condition 4 (\ref{Ultra-4}).
\end{itemize}

Thus for all horizons 
\begin{equation}  \label{G0022-N2}
\tilde{G}^0_0 -\tilde{G}^2_2 =O(N^2).
\end{equation}

Likewise the off-diagonal components in terms of $r$ can be put down in the
form 
\begin{align}
\frac{1}{\sqrt{g_{\phi \phi }}}\;\tilde{G}_{1}^{0}& =-\frac{1}{4g_{zz}}\Big\{%
\partial _{z}(\gamma _{z}-3\gamma _{\phi })\cdot \partial _{z}\omega +\frac{%
\partial _{z}N^{2}}{N^{2}}\partial _{z}\omega -2\partial _{z}^{2}\omega %
\Big\}\frac{1}{N}+O(N);  \label{G12fin} \\
\sqrt{g_{zz}}\;\tilde{G}_{2}^{3}& =
	\Big\{\frac{g_{\phi \phi }}{2}
	\partial _{r}\omega \cdot \partial _{z}\omega
	-N\partial _{r}\partial _{z}N\Big\}
	\frac{\sqrt{A}}{N^2}+O(N).
\end{align}%
In the non-extremal and extremal cases the terms in braces are $O(N^2)$ due to conditions (\ref{Non-Extra-Gen}), and in the ultra-extremal case the
same result follows from conditions 1, 3 and 5 (\ref{Ultra-1},\ref{Ultra-3},%
\ref{Ultra-5}). Note that the finiteness of scalar invariants only, which
leads to $\partial _{z}\omega =O(N)$, is not sufficient for $\tilde{G}%
_{1}^{0}$ to turn to zero; the converse result was mistakenly obtained in 
\cite{v2} due to the use of the tetrad which is not exactly orthonormal.

As the result, for all horizons 
\begin{equation}  \label{G01-23}
\tilde{G}^0_1,\tilde{G}^2_3 =O(N).
\end{equation}

\subsection{Kinematic origin of the on-horizon structure of the Einstein tensor}

Let us, following our general logic, assume that all components of the curvature tensor in the FZAMO frame $\{e_{\mu}\}$ are finite. Both groups of observers -- FZAMOs and OZAMOs -- are related by local Lorentz boosts (\ref{tetrad-fin}). Now, we will show that these two circumstances entail a rather general form of the constraints on the structure of the Einstein tensor in the OZAMO frame $\{h_{\mu}\}$ (\ref{OZAMOframe}). For simplicity, we will restrict our consideration here to FZAMOs falling relative to OZAMOs exactly  in the radial direction, so that $\theta=0$ and the second step (\ref{tetrad-mid}) in building the FZAMO tetrad is omitted. Recall that tildes denote tensors calculated in the OZAMO frame, and hats -- those in FZAMO frame.

Indeed, let us write down the boost (\ref{tetrad-fin}) in the form
\begin{equation}
e_{(\xi)}={\mathrm{x}^{\eta}}_{\xi}h_{(\eta)},\qquad 
\xi,\eta =0,2,
\end{equation}
where 
\begin{equation}{\mathrm{x}^{\eta}}_{\xi}
=\gamma\bigg(\begin{array}{cc}1&v\\ v&1\end{array}\bigg)
\end{equation}
and the Lorentz factor is
\begin{equation}\label{gammaf}
\gamma\equiv\frac{1}{\sqrt{1-v^{2}}}=\frac{E}{N},
\end{equation}
which follows from the conservation of energy
and the fact that the conserved angular momenta of both observers are equal to zero.

Then, we can calculate $\hat{G}_{00}$, $\hat{G}_{22}$ in terms of the OZAMO frame and take the horizon limit, in which $N\rightarrow0$, $v\rightarrow1$. After some elementary algebra, one finds that
\begin{align}
\hat{G}_{00}
	&=\tilde{G}_{\xi\eta}{\mathrm{x}^{\xi}}_{0}{\mathrm{x}^{\eta}}_{0}
	=\gamma^{2}\tilde{G}_{00}
	+2\gamma v\tilde{G}_{02}
	+\gamma^{2}v^{2}\tilde{G}_{22}	,\\
\hat{G}_{22}&=\tilde{G}_{\xi\eta}{\mathrm{x}^{\xi}}_{2}{\mathrm{x}^{\eta}}_{2}
	=\gamma^{2}\tilde{G}_{22}
	+2\gamma v \tilde{G}_{02}
	+\gamma^{2}v^{2}\tilde{G}_{00}.
\end{align}
In the horizon limit then, taking into account that due to symmetry (see Eq. (\ref{Gij-zeros}) in Appendix) $\tilde{G}_{02}=0$
\begin{equation}
(\hat{G}^{2}_{2}-\hat{G}^{0}_{0})\equiv
(\hat{G}_{22}+\hat{G}_{00})
	\approx 2\gamma^{2}
	(\tilde{G}_{22}+\tilde{G}_{00}),
\end{equation}
so, the finiteness of $(\hat{G}^{0}_{0}-\hat{G}^{2}_{2})$ requires that
\begin{equation}
(\tilde{G}^{0}_{0}-\tilde{G}^{2}_{2})
	=O(\gamma^{-2})=O(N^{2}).
\end{equation}

In a similar way one can easily show that
\begin{equation}
\hat{G}_{01}=\gamma\tilde{G}_{01},\quad
\hat{G}_{23}=\gamma\tilde{G}_{23},
\end{equation}
so the finiteness of $\hat{G}_{\mu\nu}$ leads also to
\begin{equation}
\tilde{G}_{01}=O(N)
\quad\text{and}\quad
\tilde{G}_{32}=O(N).
\end{equation}

The results coincide with those obtained above by direct calculations in the OZAMO frame.

\section{Summary and conclusion}

We investigated the restrictions imposed on the metric by the conditions of regularity of increasing strength: boundedness of the Ricci scalar, boundedness of quadratic scalar invariants, boundedness of tetrad components 
of the curvature tensor in a frame attached to a falling observer. The results apply to generic dirty axially symmetric rotating black holes.

Starting with the non-extremal metric written in terms of proper distance to
the horizon $n$, we saw in particular, that the regularity conditions demand
the metric to be expanded in terms of the quasiglobal coordinate $r\sim
n^{2} $ rather than $n$: for scalar invariants and tetrad components of the
curvature tensor to be bounded it is necessary that 
\begin{equation*}
g_{\mu\nu}=g_{\mu\nu}^{(h)} +g_{\mu\nu}^{(1)} r +g_{\mu\nu}^{(2)}
r^{2}+o(r^2)
\end{equation*}

For extremal and ultra-extremal metrics we wrote the expansions of metric
functions in terms of $r$ from the very beginning. The conditions of
regularity obtained in all cases, assuming the expansions in terms of integer powers of $r$, are collected in two tables: \ref{Table-Results2} in
terms of asymptotic behavior of metric functions and in \ref{Table-Results3}
in terms of coefficients of their expansions by $r$ (\ref{Ultra-metric}).

\begin{table}[th]
\centering
\begin{tabular}{|l||c|c|c|}
\hline
What is bounded & Non-extr., $p=1$ & Extr., $p=2$ & Ultra-extr., $p>2$ \\ 
\hline\hline
Ricci scalar & $\partial_{z}\omega=O(N^2)$ & \multicolumn{2}{|c|}{$%
\partial_{z}\omega=O(N)$} \\ \hline
Quadratic invariants & $\kappa=const$ & \multicolumn{2}{|c|}{$\kappa=0^{*}$}
\\ \hline
\begin{tabular}{l}
Curvature tensor \\ 
in FZAMO frame%
\end{tabular}
& \multicolumn{2}{|c|}{$\partial_{z}\ln N^{2},\partial_{z}\omega=O(N^2)$} & 
six conditions \\ \hline
\end{tabular}%
\caption{The restrictions on metric for different types of horizons that
follow from conditions of regularity of increasing strength; each line gives
the conditions for the corresponding quantity to be bounded, additional to
those stated in all the lines above (${}^{\ast }$: holds by definition).}
\label{Table-Results2}
\end{table}

\begin{table}[th]
\centering
\begin{tabular}{|l||c|c|c|}
\hline
What is bounded & Non-extr., $p=1$ & Extr., $p=2$ & Ultra-extr., $p=3$ \\ 
\hline\hline
Ricci scalar & $\omega_{H}^{\prime}=0$ & $\omega_{H}^{\prime}=0$ & $%
\omega_{H}^{\prime}=\omega_{1}^{\prime}=0$ \\ \hline
Quadratic invariants & $\kappa^{\prime}=0$ & --- & --- \\ \hline
\begin{tabular}{l}
Curvature tensor \\ 
in FZAMO frame%
\end{tabular}
& --- & $%
\begin{array}{c}
\omega_{1}^{\prime}=0, \\ 
\;\kappa_{2}^{\prime}=\kappa_{3}^{\prime}=0\;%
\end{array}%
$ & $%
\begin{array}{c}
\omega_{2}^{\prime}=0,\quad \kappa_{3}^{\prime}=0; \\ 
\kappa_{4}^{\prime}=\kappa_{5}^{\prime}=0\;\text{OR 2 exotic variants}%
\end{array}%
$ \\ \hline
\end{tabular}%
\caption{The same restrictions as in table \protect\ref{Table-Results2} in
terms of expansion coefficients of the metric functions in powers of the
quasiglobal coordinate $r$ (\protect\ref{Ultra-metric}) and for $p=3$ as an
example of ultra-extremal metric.}
\label{Table-Results3}
\end{table}

In the extremal case those imply the uniformity of the asymptotic behavior
of the lapse function $N$ and parameter $\omega $ near the horizon, that
generalize the weakest but most obvious restrictions on the surface gravity $\kappa =const$ and angular velocity of rotation $\omega _{H}=const$ for
non-extremal horizons. In the ultra-extremal case the same can be said in
loose terms, however, the explicit form of conditions (\ref{Ultra-1}--\ref{Ultra-6}) is more complicated and cannot be unambiguously interpreted in such a way.

If we write out explicitly the expansions for the lapse function and the coefficient $\omega$ in terms of $r$ for different kinds of horizons, the results read simply as follows:
\begin{align}
p=1:\quad N^{2}&=\underline{\kappa}^{2} r
	+\kappa_{2}(z)r^{2}+o(r^{2});\\
	\omega &=\underline{\omega_{H}}
	+\omega_{1}(z)r+o(r) \label{n2};\\
p=2:\quad N^{2}&=\underline{\kappa_{2}}r^{2}
	+\underline{\kappa_{3}}r^{3}
	+\kappa_{4}(z)r^{4}+O(r^{5});  \label{e1}\\
	\omega&=\underline{\omega_{H}}
	+\underline{\omega_{1}}r
	+\omega_{2}(z)r+O(r^2) \label{e2};\\
p=3:\quad N^{2}&=\underline{\kappa_{3}}r^{3} +\underline{\kappa_{4}}r^{4} +
\underline{\kappa_{5}}r^{5} +\kappa_{6}(z)r^{6}+o(r^7); \\
\omega&=\underline{\omega_{H}} +\underline{\omega_{1}}r +\underline{
\omega_{2}}r^{2} +\omega_{3}(z)r^{3}+o(r^3),
\end{align}
however for ultraextremal case ($p\geq 3$) exotic variants are also possible, given in Sec. IV B 5. The coefficients that are constants are underlined.

It is instructive to compare the relationships between aforementioned properties, analyticity and possibility to cross the horizon in the spherically symmetric space-times and in the present case. For a spherically symmetric black hole the analyticity of the metric near the horizon guarantees that an observer falling into the black hole does not experience infinite tidal forces \cite{prd08}. In the case of rotating black holes the situation is more subtle. The metric depends on two spatial variables. In the vicinity of the horizon, this dependence drops out from the asymptotic expressions in the main approximation and, moreover, within this approximation the metric can be analytical with respect to the quasiglobal coordinate. Nonetheless, the presence of the dependence on $z$ in the next corrections can give rise to infinite components of the curvature tensors for FZAMO for (ultra)extremal horizons. Say, let in (\ref{e1}) the coefficient $\kappa _{2}$ be still constant but $\kappa_{3}=\kappa _{3}(z)$. Then, in the main approximation the lapse function
looks analytical in term of $r$ but infinite curvature components arise due to the term with $\kappa_{3}$. Therefore, it turned out that such fundamental conditions like rigid rotation and analiticity of the metric in
the immediate vicinity of the horizon are necessary but, in contrast to non-extremal ones, not sufficient for the metric near the (ultra)extremal horizon to be completely regular.

It was shown in \cite{v2} that in the vicinity of a non-extremal horizon the Einstein tensor in the (orbital) ZAMO frame diagonalizes. We generalized this result to horizons of all degrees of extremality and showed that if the tetrad components of the curvature tensor in the frame attached to a falling observer stay finite, then 
\begin{equation*}
\tilde{G}^{0}_{0}-\tilde{G}^{2}_{2}=O(N^2),\qquad \tilde{G}^{0}_{1},\tilde{G}%
^{2}_{3}=O(N).
\end{equation*}

\begin{acknowledgments}
The work of O. Z. was supported in part by the Cosmomicrophysics section of
the Programme of the Space Research of the National Academy of Sciences of
Ukraine.
\end{acknowledgments}

\section{Appendix: Einstein tensor in ZAMO frame}

The explicit expressions for the Einstein tensor in the OZAMO frame (\ref%
{OZAMOframe}), denoted by tildes, in terms of the proper distance to the
horizon $n$ are as follows. The diagonal ones are 
\begin{align}
\tilde{G}^{0}_{0}&= \frac{g_{\phi\phi}}{4}\Big[ \Big(\frac{\partial_{n}\omega%
}{N}\Big)^{2} +\!\frac{1}{g_{zz}} \Big(\frac{\partial_{z}\omega}{N}\Big)^{2}%
\Big] +\frac{1}{2}\Big\{D^{2}_{n}g_{zz} +D^{2}_{n}g_{\phi\phi} +\tfrac{1}{%
g_{zz}}D^{2}_{z}g_{\phi\phi} \Big\} +  \notag \\
&+\frac{1}{4}\big[ \partial_n \gamma_\phi\,\partial_n \gamma_z -\tfrac{1}{%
g_{zz}} \partial_z \gamma_\phi\,\partial_z \gamma_z \big]; \\
\tilde{G}^{2}_{2}&= \frac{g_{\phi\phi}}{4}\Big[ \Big(\frac{\partial_{n}\omega%
}{N}\Big)^{2} -\frac{1}{g_{zz}} \Big(\frac{\partial_{zz}\omega}{N}\Big)^{2}%
\Big] +\frac{1}{2}\Big\{ \tfrac{1}{g_{zz}}D^{2}_{z}g_{\phi\phi} \Big\} + 
\notag \\
&+\frac{1}{4}\big[ \partial_n \gamma_\phi\,\partial_n \gamma_z -\tfrac{1}{g_z%
} \partial_z \gamma_\phi\,\partial_z \gamma_z \big] +\frac{1}{2}\frac{%
\partial_n N}{N} \partial_{n}(\gamma_\phi +\gamma_z ) +\frac{1}{2g_{zz}}%
\frac{\partial_z N}{N} \partial_{z}(\gamma_\phi -\gamma_z ) +\Big[\frac{1}{%
g_{zz}}\frac{\partial_{z}^{2}N}{N}\Big]; \\
\tilde{G}^{1}_{1}&= -\frac{3 g_{\phi\phi}}{4}\Big[ \Big(\frac{%
\partial_{n}\omega}{N}\Big)^{2} +\frac{1}{g_{zz}} \Big(\frac{%
\partial_{z}\omega}{N}\Big)^{2}\Big] +\frac{1}{2}\Big\{D^{2}_{n}g_{zz} \Big\}%
+  \notag \\
&\qquad\qquad\qquad\qquad\qquad \qquad\quad +\frac{1}{2}\frac{\partial_n N}{N%
} \partial_{n}\gamma_z -\frac{1}{2g_{zz}}\frac{\partial_z N}{N}
\partial_{z}\gamma_z +\Big[\frac{\partial_{n}^{2}N}{N}+ \frac{1}{g_{zz}}%
\frac{\partial_{z}^{2}N}{N}\Big]; \\
\tilde{G}^{3}_{3}&= -\frac{g_{\phi\phi}}{4}\Big[ \Big(\frac{%
\partial_{n}\omega}{N}\Big)^{2} -\frac{1}{g_{zz}} \Big(\frac{%
\partial_{z}\omega}{N}\Big)^{2}\Big] +\frac{1}{2}\Big\{D^{2}_{n}g_{\phi\phi} %
\Big\}+  \notag \\
&\qquad\qquad\qquad\qquad\qquad \qquad\quad +\frac{1}{2}\frac{\partial_n N}{N%
} \partial_{n}\gamma_\phi +\frac{1}{2g_{zz}}\frac{\partial_z N}{N}
\partial_{z}\gamma_\phi +\Big[\frac{\partial_{n}^{2}N}{N}\Big],
\end{align}
where the notation is used 
\begin{equation}  \label{D2}
D^{2}_{a}g_{\xi}=\frac{1}{\sqrt{g_{\xi}}}\partial_{a} \frac{%
\partial_{a}g_{\xi}}{\sqrt{g_{\xi}}}, \qquad a=n,z,\quad \xi=\phi\phi,zz;
\end{equation}
the off-diagonal components 
\begin{align}
\frac{1}{\sqrt{g_{\phi\phi}}}\tilde{G}^{0}_{1}=& \frac{1}{2}\Big[ \frac{%
\partial^2_n \omega}{N} +\frac{1}{g_z}\frac{\partial_z^2 \omega}{N}\Big] -%
\frac{1}{2}\Big[ \frac{\partial_n N}{N}\frac{\partial_n \omega}{N} +\frac{1}{%
g_z} \frac{\partial_z N}{N}\frac{\partial_z \omega}{N}\Big]  \notag \\
&\qquad\qquad +\frac{1}{4}\Big[\frac{\partial_n \omega}{N}
\partial_{n}(3\gamma_\phi +\gamma_z) +\frac{1}{g_{zz}}\frac{\partial_z \omega%
}{N} \partial_{z}(3\gamma_\phi -\gamma_z)\Big]; \\
\sqrt{g_{zz}}\tilde{G}^{2}_{3}=& \frac{g_{\phi\phi}}{2} \frac{\partial_n
\omega}{N} \frac{\partial_z \omega}{N} -\frac{\partial_n \partial_z N}{N} -%
\frac{1}{2}\frac{\partial_n \partial_z g_{\phi\phi}}{g_{\phi\phi}} +\frac{1}{%
2}\frac{\partial_z N}{N} \partial_n \gamma_z +\frac{1}{4g_{zz}} \partial_z
\gamma_\phi\; \partial_n (\gamma_\phi +\gamma_z); \\
&\tilde G^{0}_{2}=\tilde G^{0}_{3} =\tilde G^{1}_{2}=\tilde G^{1}_{3}=0. \label{Gij-zeros}
\end{align}

\end{document}